\documentstyle[aps]{revtex}

\begin{document} \input psfig.sty  

\title{Phase Transitions Driven by Vortices in 2D Superfluids and 
Superconductors: From Kosterlitz-Thouless to 1st Order}

\author{G. Alvarez and H. Fort \\
Instituto de F\'{\i}sica, Facultad de Ciencias, \\ Igu\'a 4225, 11400
Montevideo, Uruguay}

\maketitle

\begin{abstract}

The Landau-Ginzburg-Wilson hamiltonian is studied for different
values of the parameter $\lambda$ which multiplies the quartic 
term (it turns out that this is equivalent to 
consider different values of the 
coherence length $\xi$ in units of the lattice spacing $a$).
It is observed that amplitude fluctuations can change dramatically the 
nature of the phase transition: for small  
values of $\lambda$ ($\xi/a > 0.7$)
, instead of the smooth Kosterlitz-Thouless
transition there is a {\em first order} transition
with a discontinuous jump in the vortex density $v$ and a larger
non-universal drop in the helicity modulus.
In particular, for $\lambda$ sufficiently small ($\xi/a \cong 1$) 
, the density of bound pairs of vortex-antivortex below $T_c$
is so low that, $v$ drops to zero almost for all temperature
$T<Tc$. 

\end{abstract}

\vspace{2mm}

Vortices play a central role in explaining the phase diagram 
and properties of superfluid systems both neutral and charged. 
The discovery
of high-temperature superconductors has boosted the interest in
the dynamics of the vortex lines in the mixed state and opened
a new area which regards the physics of vortices as a new state
of matter \cite{bl-fe-ge-la-vi94}. 

Superfluid films and Josephson junction arrays 
in two dimensions are often described by XY-type models in which the
unique microscopic degrees of freedom are phases. 
Thanks to the work of Berezinskii \cite{be72} and Kosterlitz and Thouless 
\cite{ko-th73} in the early 70's we have a fair understanding of the
{\em standard} XY model in two dimensions. In their theory a thermodynamic
phase transition, the Kosterlitz-Thouless (KT) transition is driven by
the unbinding vortices ( singular phase configurations )
at a temperature $T_{KT}$.
In refs. \cite{be-mo-or79}-\cite{he-fi83} it was pointed out that the 
KT phase transition might also apply to thin-film superconductors and some
experimental evidence was discussed. In particular, the 
analysis of the 2 dimensional flux line lattice (FLL) melting transition 
of ref. \cite{do-hu79} is based on a KT-type theory. 
More recently, it was suggested that some features of the
KT theory might be present in under-doped superconducting cuprates 
\cite{nature99}. 

However, the nature of the phase transition driven by vortices in 2D
still remains under discussion. By means of Monte Carlo simulations
of the XY model with a modified nearest neighbor interaction it was
shown that, depending on the value of an additional parameter, 
continuous as well first-order transitions take place 
\cite{jo-mi-ny93}-\cite{hi84}. 
The existence of both kinds of phase transitions is 
in accordance with the richer structure of 
the 2D Coulomb gas found by Minnhagen and Wallin \cite{mi-wa87} using 
self-consistent renormalization group equations and with the tendency
towards first-order transition which develops in the case of
a strong disorder coupling constant \cite{ko92}.

Our approach to the nature of the phase transition
driven by vortices in 2D superfluids/superconductors is 
a different one based on the effect of amplitude fluctuations. 
Let us start by noting
that right at the ( expected vortex-pair unbinding )
transition the amplitude fluctuations cannot be considered as weak 
anymore and thus may affect the critical behavior \cite{bo-be94}. 
Therefore, instead of the XY model -with fixed amplitude fields-
it is worth investigating the effects of vortices when amplitude 
fluctuations are not neglected. Hence, in this letter
we analyze the more general 
{\em Landau-Ginzburg-Wilson} (LGW) \cite{ba80} lattice
Hamiltonian, in terms of a complex field $\psi=|\psi|\exp{i\theta}$
($|\psi| \not=$ constant) and two parameters $K$ and $\lambda$ 
\cite{parameterization}:
\begin{equation} 
\beta H = -2K \sum_x \sum_\mu |\psi_x||\psi_{x+a\mu}| cos(\theta_{x+a\mu}-\theta_x)
+ \sum_x [ \lambda ( |\psi_x|^2 -1 )^2 + |\psi_x|^2],
\label{eq:H1}
\end{equation} 
where $\beta=1/T$, $a$ is the lattice spacing, $x$ denotes the lattice sites and
the index $\mu=1,2$. 
We will show that the nature of the phase transition of this model 
diverges dramatically from the KT when the parameter $\lambda$ is chosen 
sufficiently small -which in fact is equivalent to take the coherence 
length $\xi \simeq a$- and that this is connected with the appearance of a 
sharp jump in the number of vortices.

A straightforward discretization of the continuum Landau-Ginzburg 
Hamiltonian produces the expression  \cite{ba80}: 
\begin{equation} 
\beta H = \beta a^2 \sum_x [ \sum^2_{\mu=1} \frac{\hbar^2}{2m} (
\psi^c_{x+a\mu}-\psi^c_x)^2/a^2 +r |\psi^c_x|^2 + u |\psi^c_x|^4 ],
\label{eq:H2}
\end{equation}
where the superscript $c$ in $\psi$ denotes the ordinary parameterization
in the continuum theory, 
$m$ is the effective mass of the carriers
and the coefficients $r$ and $u$
are analytic functions of the temperature, with $u >0$ for stability. 
Introducing a dimensionless order parameter: $\bar{\psi}_x =
\left(\frac{2u}{r}\right)^{\frac{1}{2}}\psi^c_x$ and writing
\begin{equation} 
\beta H = \frac{1}{k_BT} a^2 \sum_x [ \sum^2_{\mu=1}  
(\bar{\psi}_{x+a\mu}-\bar{\psi}_x)^2/a^2 + \frac{1}{2\xi^2}
(1-|\bar{\psi}_x|^2)^2 ],
\label{eq:H3}
\end{equation}
where $\frac{1}{k_BT}=\frac{\hbar^2|r|}{2m u}\beta$ and $\xi$ is the 
coherence length given by $\xi^2=
\frac{\hbar^2}{2m |r|}$. Parameterizations (\ref{eq:H1}) and (\ref{eq:H3}) 
are connected by the relations:
\begin{equation}
\frac{\xi}{a}=
\left(\frac{K}{|1-2\lambda-4K|}\right)^\frac{1}{2},
\;\;\;\;\;\;\;\;\;
T=\frac{\lambda (\xi/a)^2}{K^2}=\frac{\lambda}{K|4K+2\lambda-1|}.
\label{eq:connec}
\end{equation}

In the limit of $\lambda=\infty$ the radial degree of freedom is 
frozen and this model -sometimes said to describe
{\em soft spins} with non fixed amplitude- becomes the XY 
model which is said to describe {\em hard spins} with fixed amplitude. 
A more interesting and less well studied 
limit is just the opposite i.e. small values of the $\lambda$
parameter. By (\ref{eq:connec}) 
small values of $\lambda$ correspond to a large $\xi$ in units of 
$a$ i.e. $\lambda$ and 
$1/\xi^2$ are the self-interaction coefficients respectively in 
(\ref{eq:H1}) and (\ref{eq:H3}) regulating amplitude fluctuations.

We have simulated the Hamiltonian (\ref{eq:H1}) using 
a Monte Carlo algorithm. The calculations were performed
on square $L\times L$ lattices  
with periodic boundary conditions (PBC).
In order to increase the speed of the simulation we have discretized
the $O(2)$ global symmetry group to a $Z(N)$ and compared the results with
previous runs carried out with the full O(2) group in relatively small 
lattices. For the case of $Z(60)$ we found no appreciable differences.
Lattices with $L=10, 20, 24, 32, 40$ and (in some cases) 64 were used. 
For  
$L=10, 20, 24$ we thermalized with, usually, 20.000-40.000 sweeps and
averaged over another 60.000-100.000 sweeps. For $L=32, 40$ and 64  
larger runs
were performed, typically 50.000 sweeps were discarded for 
equilibration and averaged over 200.000 sweeps. We also performed
some more extensive runs near $K_c$ for small values of $\lambda$.  
The following quantities were measured: i) The {\em vortex density v}.
The standard procedure to calculate the vorticity  
on each plaquette is by considering the quantity
\begin{equation}
m=\frac{1}{2\pi}([{\theta}_1 - {\theta}_2 ]_{2\pi} +
[{\theta}_2 - {\theta}_3 ]_{2\pi} +
[{\theta}_3 - {\theta}_4 ]_{2\pi} +
[{\theta}_4 - {\theta}_1 ]_{2\pi}),
\label{eq:m}
\end{equation}
where $[{\alpha}]_{2\pi}$ stands for $\alpha$ modulo 2$\pi$:
$[{\alpha}]_{2\pi} = \alpha + 2{\pi}n$, with $n$ an integer
such that $\alpha + 2{\pi}n
\in (-\pi ,\pi ]$, hence
$m=n_{12}+n_{23}+n_{34}+n_{41}$.
%\label{eq:j1}
%\end{equation} 
If $m\neq 0$, there exists a {\em vortex} which is assigned to the
object dual to the given plaquette. Hence in the case d = 2, $*m$, the dual
of $m$, is assigned to the center of the original plaquette $p$.
The vortex ``charge" $*m$ can take three values: 0, $\pm 1$ (the value 
$\pm 2$ has a negligible probability).
$v$ defined as:
\begin{equation}
v=\frac{1}{L^2} \sum_{x} |*m_x|,
\label{eq:v}
\end{equation}
serves as a measure of the vortex density.
ii) The {\em energy density} $\varepsilon=<H>/L^2$ and the 
{\em specific heat} $C_v$, which were
computed to measure the order of the phase transition.
iii) The {\em helicity modulus} $\Gamma$ \cite{fi-ba-ja73}
which measures the phase-stiffness. For a spin system with PBC 
the helicity modulus measures the cost in free energy of imposing a ``twist" 
equal to $L\delta$ in the phase between two opposite boundaries of the system.
$\Gamma$ is obtained in general as a second order derivative of 
the free energy with respect to $\delta$ -which can be regarded as a uniform
statistical vector potential- evaluated for $\delta \rightarrow 0$.
In such a way one gets the following expression \cite{eb-st83},
%\cite{sh-eb-st84}
which generalizes the one introduced in ref. \cite{fi-ba-ja73}, to an order
parameter with amplitude as well as phase variations:
\begin{equation}
\Gamma =\frac 1N \{ <\sum_{<ij>}~' \mid \psi_i\mid \mid \psi_j \mid \cos
(\theta_i - \theta_j)> - k<{[}\sum_{<ij>}~'\mid \psi_i \mid \mid \psi_j\mid \sin
(\theta_i - \theta_j){]}^2>\},
\label{eq:helicity}
\end{equation}
where the primes denote that the sums are carried out over links
along one of the 2 directions (x or y).

\begin{center}
\begin{figure}[h]
\centering
\psfig{figure=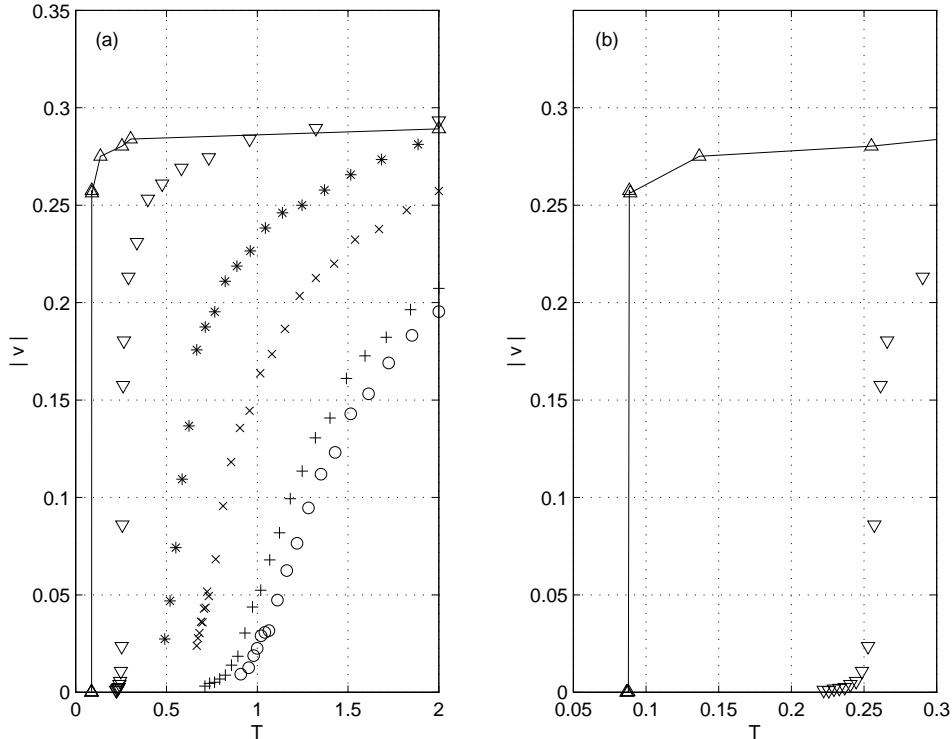,height=10cm}
\caption{(a): $v$ vs. $T$ for $\lambda=0.01$ (triangles up), 
$\lambda=0.1$ (triangles down),
$\lambda=1$ (*), $\lambda=2.5$ ($\times$), $\lambda=10$ (+)
and XY (circles).
(b): zoom of 1-(a) to compare the drop to 0 of $v$
for $\lambda=0.01$ with the less sharp drop 
for $\lambda=0.1$.}
\end{figure}
\end{center}

Fig. 1-(a) shows a plot of $v$ vs. $T$ for different values of $\lambda$
and $L=40$.
For $\lambda=0.01$ ($\frac{\xi}{a}\cong 0.94$ 
we observe a sharp jump in the vortex density $v$
(triangles up). As long as we
increase $\lambda$ the jump becomes more smooth and moves to higher values of 
$T_c$ until for $\lambda=10$ (+ symbols) we get something very close
to the $\lambda \rightarrow \infty$ KT behavior (circles).
The increase in the density of vortices when amplitude fluctuations are
large is in agreement with the analytical computations 
of ref. \cite{bo-be94}. What is new is the fact that, 
when $\xi\simeq a$  
($\lambda \cong 0.01$), the transition occurs almost directly from 0 vortex 
to a plasma of vortices i.e. the bound pairs of
vortex-antivortex seem to play no major role. Fig. 1-(b) is a zoom 
of 1-(a) showing the difference between $\lambda=0.01$ and $\lambda=0.1$
below $T_c$: for $\lambda=0.1$ $v$ drops to non-zero values due
to the existence of vortex-antivortex pairs while for $\lambda=0.01$
$v$ drops much more sharply to 0 signaling the sudden extinction of vortices.

The different behavior of $v$ above $T_c$
between the large fluctuating amplitude 
{\em LGW-regime} $\frac{\xi}{a} \sim 1$ and the {\em KT-regime}  
$\frac{\xi}{a} \ll 1$ is because
amplitude fluctuations -governed by the $\lambda$ parameter- indeed
decrease the energy of vortices enhancing vortex production.
The same happens for the XY model with modified interaction \cite{hi84};
in fact, the shape modification can be 
straightforwardly connected to a core energy variation.

The scarcity of bound pairs of vortex-antivortex below Tc for the 
extreme fluctuations regime
(very small values of $\lambda$ or $\xi\sim a$) can be explained in terms
of the behavior of their free energy  
$F_{pair}=E_{pair}-TS_{pair}$ at $T \sim T_c$ from below. 
Roughly, $E_{pair} \simeq 2E_c$, where $E_c$ 
is the vortex core energy, and  
$S_{pair} \sim \ln (\frac{L^2}{\xi^2})$. The three quantities
$S_{pair}, E_c$ and $T_c$ all decrease
as $\lambda$ decreases making difficult to 
disentangle the "energetic" contribution to $v$, proportional to 
$\exp[-E_{pair}/T_c]$, from the "entropic" contribution $\sim 1/\xi^2$.
For the intermediate range of $\lambda$ (or $\xi$), although it is not easy to
predict the behavior of the energetic factor, the entropy seems to be
the main responsible for lowering the density of bound 
pairs \cite{fit}. 
From a $\lambda$ sufficiently small, $T_c$ starts to decrease with $\lambda$
faster than $E_{pair}$ and thus 
$\exp[-F_{pair}/T]$ decreases more and more sharply making
smaller and smaller the probability of bound pairs of vortex-antivortex.

Figures 2-4 show plots of $\varepsilon$, $v$ and $\Gamma$ 
for $\lambda=0.01, 0.1$ and 10. 
For $\lambda = 0.01$ the transition is clearly first-order:
we observe latent heat and 
discontinuous changes in $v$ and $\Gamma$ at $T=T_c$. 
On the other hand, 
for $\lambda=10$ (or $\frac{\xi}{a} \simeq \frac{1}{\sqrt{42}}$ )
the results are similar to those of the XY model. In particular,
we get something close to the KT universal jump  $\Delta \Gamma =
\frac{2}{\pi}$. As long as
$\lambda$ decreases we get a larger {\em non universal} jump in 
$\Gamma$. $\lambda=0.1$ corresponds to something in between 
first order and KT.

\begin{center}
\begin{figure}[h]
\psfig{figure=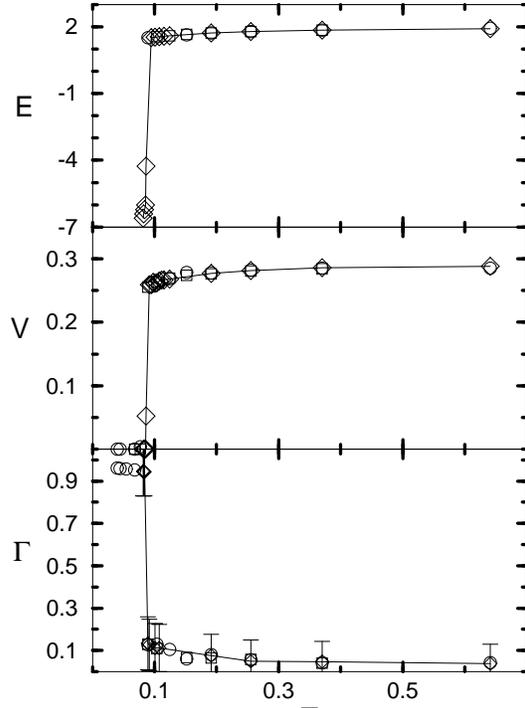,height=9cm} 
\caption{$\varepsilon$, $v$ and $\Gamma$ vs. T , $\lambda=0.01$ for sizes:
$L=10$ (circles), $L=20$ (diamonds) and $L=40$ (squares). Error bars for
$\varepsilon$ and $v$ are smaller than the symbol sizes.}
\end{figure}
\end{center}

\begin{center}
\begin{figure}[b]
\psfig{figure=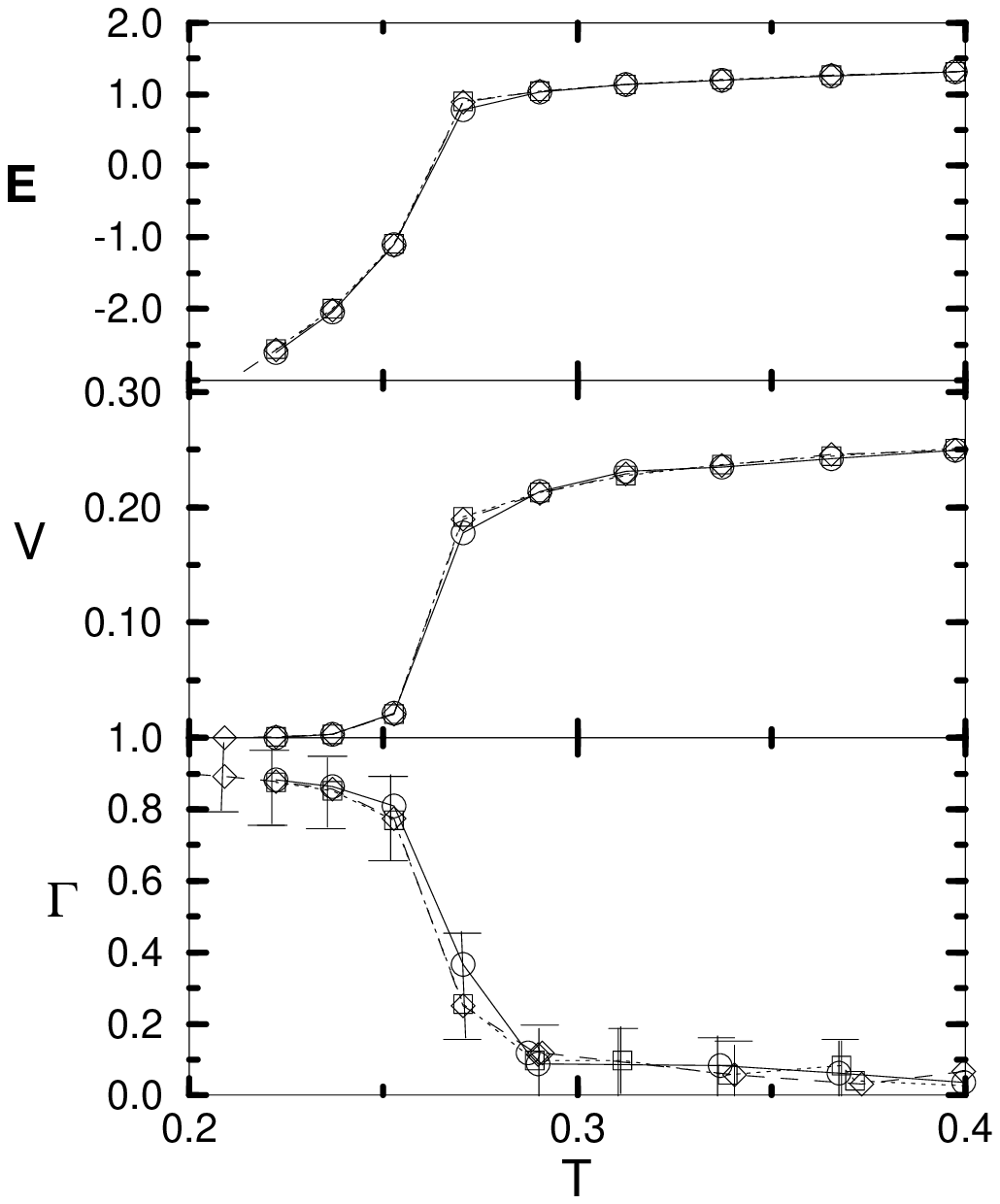,height=9cm} 
\caption{$\varepsilon$, $v$ and $\Gamma$ vs. T, $\lambda=0.1$ for sizes:
$L=10$ (circles), $L=20$ (diamonds) and $L=40$ (squares). Error bars for
$\varepsilon$ and $v$ are smaller than the symbol sizes.}
\end{figure}
\end{center}

\begin{center}
\begin{figure}[h]
\psfig{figure=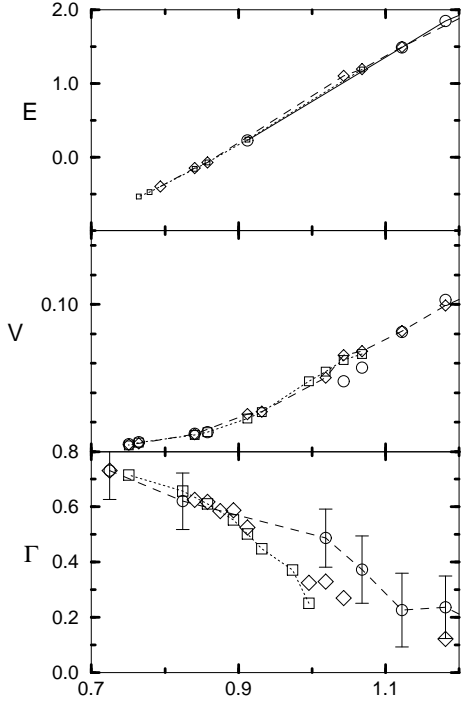,height=9cm} 
\caption{ $\varepsilon$, $v$ and $\Gamma$ vs. T, $\lambda=10$ for sizes: 
$L=10$ (circles), $L=20$ (diamonds) and $L=40$ (squares). Error bars for
$\varepsilon$ and $v$ are smaller than the symbol sizes.}
\end{figure}
\end{center}
The double peak structure corresponding to the 2 coexisting phases, 
characteristic of a first-order transition, is showed in Fig. 5(a) for
$\lambda=0.01$ and sizes $L=10,20$ and 40. The width $w$ of each
of the peaks clearly scale as $w\sim \sqrt(\frac{1}{L^D})=\frac{1}{L}$ 
due to ordinary non-critical fluctuations. Fig 5(b) is a zoom
of the right peak centered around the higher energy phase. It shows
that $w \sim 0.1 (L=40), w \sim 0.2 (L=20)$ and $w \sim 0.4 (L=10)$.

Therefore, one has a simple model in which the nature of the phase 
transition depends on the value of one parameter ($\lambda$ or
$\xi/a$)
which controls the thermal fluctuations of vortex cores and
from which the following picture emerges:

1) For $\lambda < 0.1$ or $\frac{\xi}{a} \sim 1$
the density of vortices experiments an abrupt jump which coincides with a 
first order transition with large latent heat and a {\em non universal}
jump in $\Gamma$ all
at a $T_c$ which grows with $\lambda$. This is the 
LGW-regime as opposed to the much more smooth KT-regime.
In particular, for sufficiently small values of $\lambda$ 
(for instance $\lambda = 0.01$) below $T_c$ 
$v$ drops to zero i.e. the number of measured bound pairs of 
vortex-antivortex is negligible compared with the number
found in the KT transition.

2) For $\lambda \ge 10$ or $(\frac{\xi}{a})^2\ll 1$ we get basically the XY
model ($\mid \psi \mid $ is fixed = 1 for all K) and the more subtle 
KT transition (with an unobservable essential singularity in the specific
heat at $T_c$ and a much more small non-universal maximum above $T_c$ 
and a universal jump in $\Gamma$).
The number of vortices and the energy evolve smoothly across the transition.

3) For intermediate values of $\lambda$ ($0.1 \le \lambda < 10$) we have an
interpolating regime between LGW and KT.

\begin{center}
\begin{figure}[h]
\psfig{figure=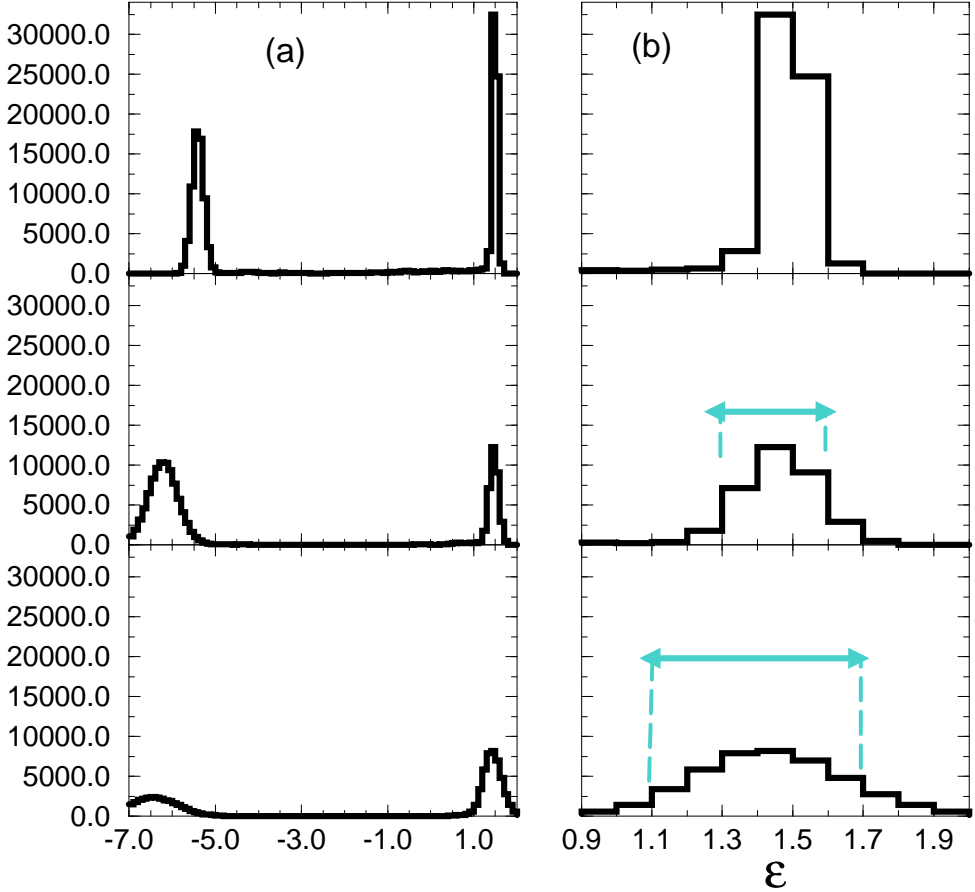,height=11cm} 
\caption{(a)Histograms of $\varepsilon$ for $\lambda=0.01$ and 
$L=10,20$ and $40$ at $T=T_c$. (b) Zoom of the right peak.}
\end{figure}
\end{center}

Whether or not a large enough increment of
$\frac{\xi}{a}$ or $\lambda$ to alter the nature of the 
phase transition driven by vortices can be accomplished 
by varying some thermodynamic parameter, for instance the pressure, 
is something which deserves investigation.

This change in the nature of the phase transition when amplitude
fluctuations of $\psi$ are not negligible ($\lambda \le 0.1$) is 
in agreement with very recent variational computations for the same model
\cite{be-cu99}. 
Furthermore, we checked that all the couples of values ($K_c,\lambda$) 
for what we found
a first order transition are such that equation 
(\ref{eq:connec}) gives $0.7 <
\xi/a < 1.1$ in complete accordance with figure 6 of ref. \cite{bo-be94}
from which we can see that for $\xi/a$ bigger than 0.5 the 
RG trajectories cross the
first order line of Minnhagen's
\cite{mi-wa87} generic phase diagram for the two-dimensional
Coulomb gas \cite{comp}. The jump in $\Gamma$ larger than the universal
value seems consistent with experiments on thin films HT$_c$ superconductors
\cite{le90}.

Finally, our analysis could shed light on the 
nature of the melting transition for the 2D FLL which remains 
controversial theoretically as well as experimentally.  
Experimental works and numerical simulations favor a KT-like 
transition in 
some cases and a discontinuous one in others \cite{st88}.
A recent extensive Monte Carlo simulation found a 
first order transition at a temperature close to the
estimated one assuming a KT melting transition \cite{ka-na93}.
After all, it is possible that the nature of the melting transition in 2D
strongly depends on the particular conditions and details of the studied 
specimen which in turn translate into different values of $\frac{\xi}{a}$.

\vspace{1mm}

Work supported in part by CSIC, Project No.  052 and 
PEDECIBA.  We are indebted with D. Ariosa, H. Beck, P. Curty, E. Dagotto 
and  G. Gonzalez-Sprinberg for valuable discussions.

%\vspace{3mm}

%Fig.2: $\varepsilon$, $v$ and $\Gamma$ vs. $T$ for $\lambda=0.01$.
%
%\vspace{3mm}
%
%Fig.3: $\varepsilon, v$ and  $\Gamma$ vs. $K$ for $\lambda=0.1$.
%
%\vspace{3mm}
%
%Fig.4: $\varepsilon, v$ and  $\Gamma$ vs. $K$ for $\lambda=10$.
%
%\vspace{3mm}
%
%
%Fig. 5: Histograms of $\varepsilon$ for $\lambda=0.01$ and $L=10,20,40$.


\begin{thebibliography}{99}
\bibitem{fit} We measured both $E_c$ and $v$ for different values of 
$\xi$ at $T \sim T_c$ from below and found that the entropic factor 
fits in well with $v$ in the intermediate range of $\xi$ while the 
energetic factor is the dominant one when $\xi \sim 1$ and $v$
becomes neglegible.
\bibitem{comp} Formulas (87) and (88) of \cite{bo-be94},
connecting ($\xi$,$T$) with ($T_{CG}$,$z_{CG}$),  provide a link
between our calculations and Minnhagen phase diagram.  However, one
should not compare directly values of $(\lambda,K)$ to $(T_{CG}$,
$z_{CG}$),
since the former are initial model parameters and the latter vary during
renormalization according to the curves shown in B-B figure 6.
\end{thebibliography}
\end{document}